\documentclass[a4paper,11pt]{article}

\usepackage{amssymb,amsmath,amsfonts}
\usepackage[left=2cm, right=2cm, top=2.5cm, bottom=2.5cm]{geometry}
\usepackage[x11names]{xcolor}
\usepackage{fancyhdr, amssymb, cancel, amsmath, graphicx, pgfplots, tikz}
\usepackage{float}
\usepackage{isomath}
\usepackage[toc]{appendix}
\usetikzlibrary{shadows}

\newcommand{\scolor}{blue!60!black}
\newcommand{\tcolor}{red!60!black}

\usepackage[colorlinks=true, urlcolor=\scolor, 
linkcolor=\tcolor, citecolor=\scolor!80!white, hyperindex=true, linktocpage=true]{hyperref}
\usepackage[explicit]{titlesec}

\def\be{\begin{equation}}
\def\ee{\end{equation}}
\def\bea{\begin{eqnarray}}
\def\eea{\end{eqnarray}}

\def\p{\pi}                
\def\a{\alpha}
\def\s{\sigma}

\renewcommand{\Im}{\mbox{Im} \;}
\renewcommand{\Re}{\mbox{Re} \;}

\def\pa{(2 \pi \alpha')}

\def\Nf{N_f}
\def\Nc{N_c}

\def\cp{\mathbb{CP}^2}
\def\nn{\nonumber}
\def\Eb{E_{\beta}}
\def\lag{\langle}
\def\rag{\rangle}

\def\Qx{\lag J^x \rag}
\def\a{\alpha}
\def\b{\beta}

\def\cp{\mathbb{CP}^2}
\def\nn{\nonumber}
\def\nn{\nonumber}
\def\Eb{E_{\beta}}
\def\lag{\langle}
\def\rag{\rangle}
\def\L{L^2}

\def\hri#1#2{\href{http://arxiv.org/abs/#1}{[ArXiv:#1]#2}}

\begin{document} 
    \fancyhead{}
    
    \fancyfoot{}
    \fancyfoot[C] {\textsf{\textbf{\thepage}}}
    \begin{equation*}
        \begin{tikzpicture}
        \draw (0.5\textwidth, -3) node[text width = \textwidth] { \huge \textsf{\textbf{ Ground state instability in Nonrelativistic QFT and\\ \vspace{0.5 cm} Euler-Heisenberg Lagrangian  via holography }} };
        \end{tikzpicture}
    \end{equation*}

\begin{equation*}
    \begin{tikzpicture}
    \draw (0.5\textwidth, 0.1) node[text width=\textwidth] {\large \color{black}  \textsf{{\color{\scolor}Ali Vahedi}  {\color{\scolor} }}};
    \draw (0.5\textwidth, -0.5) node[text width=\textwidth] {\small\textsf{Departemet of Physics, Kharazmi University, Mofatteh Ave, Tehran, Iran}};
    \end{tikzpicture}
\end{equation*}
\begin{equation*}
    \begin{tikzpicture}
    \draw (0, -13.1) node[right, text width=0.5\paperwidth] {\texttt{vahedi@khu.ac.ir , vahedi@ipm.ir }};
    \draw (\textwidth, -13.1) node[left] {\textsf{}};
    \end{tikzpicture}
\end{equation*}

\allowdisplaybreaks

\pagestyle{fancy}
\renewcommand{\headrulewidth}{0pt}
\fancyhead{}
\fancyfoot{}
\fancyfoot[C] {\textsf{\textbf{\thepage}}}

\begin{equation*}
\begin{tikzpicture}
\draw (\textwidth, 0) node[text width = \textwidth, right] {\color{white} };
\end{tikzpicture}
\end{equation*}


\begin{center}
    \begin{equation*}
    \begin{tikzpicture}
        \draw (0.12\textwidth, -6.25) node[left] {\color{\scolor}  \textsf{\textbf{Abstract:}}};
    \draw (0.53\textwidth, -6) node[below, text width=0.8\textwidth, text justified] {\small We study the ground state instability of a strongly coupled QFT with the $z=2$ Schr\"odinger symmetry in a constant electric field using probe branes holography. The system is $N_f$ $\mathcal{N}=2$  hypermultiplet fermions at zero charge density in the supergravity Schr\"odinger background. We show that the instability occurs due to Schwinger-like effect and an insulator state will undergo a transition to a conductor state. We calculate the decay rate of instability and pair production probability by using the $gauge/gravity$ duality. At zero temperature for massive fermions, we suggest that the instability occurs if the critical electric field is larger than the confining force between fermions, which is proportional to an effective mass. We demonstrate that, at zero temperature, the Schr\"odinger background simulates the role of a crystal lattice for massive particles. We also show that at finite 't Hooft coupling for particles with a mass higher than $\frac{\sqrt{\lambda}}{\pi \b}$, in this background, instability does not occur, no matter how large the external electric field is, meaning that we have a \textit{perfect insulator}. Moreover, we derive Euler-Heisenberg effective Lagrangian for the non-relativistic strongly correlated quantum theory from probe branes holography in Schr\"odinger spacetime.};
    \end{tikzpicture}
    \end{equation*}
    %
\end{center}

\newpage

\tableofcontents
\section{Introduction}
The decay of false vacuum to true vacuum might be considered as a pair production mechanism. The Schwinger pair production results an instability of the vacuum in QED ~\cite{Schwinger:1951nm,Heisenberg1936}. 
In the condensed matter systems ground state or vacuum could also experience instability in the same situation. For example, the electric breakdown of an insulator, when subjected to a high external electric field, could be considered as instability in vacuum. At the presence of an external electric field, the decay of the ground state to ground state in condensed matter physics is considered as the Zener breakdown of the Mott (or band) insulator~\cite{oka:514}. The revisited version of Euler-Heisenberg effective Lagrangian in condensed matter physics is studied through the ground state to ground state transition amplitude or the Zener tunneling rate~\cite{Oka:2011kf}.
\\
Generally, the Schwinger effect as a vacuum instability is a non-perturbative phenomenon. In strongly correlated systems, calculating the Schwinger effect demands great effort. The well-known toolbox for studying robust coupled systems in quantum field theories is $ AdS/CFT$ correspondence or more generally the $gauge/gravity$ duality. 
The original AdS/CFT correspondence states that: The the $AdS_5 \times S^5$, as the near extremal solution of $N_c$ coincident $D3-$branes, is dual to $3+1$ dimensional super-conformal field theory with $SU(N_c)$ gauge degrees of freedom. The gauge degrees are shown by the adjoint representation of the $SU(N_c)$. For adding other degrees of freedom in addition to of gauge fields, we add $N_f$ $D7-$branes, see~\cite{Karch:2002sh}, and for the simplicity, consider them as a probe ($N_f \ll N_c$). This configuration describes a QCD-like system~\cite{Karch:2002sh}. In general, this means that we add   $\mathcal{N}=2$ hypermultiplet fermions(quarks) in fundamental representation to the background gauge theory. From the gravity side, these fermions are the strings with one end on $D7$ branes and the other on $D3$ branes. Dynamics of the $D3/D7$ system is given by the $DBI$ action of probe $D7$ branes. From the gauge/gravity dictionary, this action would be an effective action of fermions in the boundary theory. The vacuum instability of the supersymmetric QED(QCD) is studied through probe brane holography in the AdS background in the~\cite{Hashimoto:2013mua}. Results are in agreement with Schwinger pair production in QED after replacing 't Hooft coupling $\lambda$ with the QED coupling constant, $e^2$. We aim to generalize this idea to QFT with the Schr\"odinger symmetry instead of the conformal symmetry. The cold atoms system known as fermions at unitarity is a famous example of a system with the Schr\"{o}dinger symmetry studied using holography in the~\cite{Son:2008ye}, see also~\cite{Taylor:2008tg,Andrade:2014iia,Andrade:2014kba,Guica:2010sw}.
\\
By using null Melvin twist($NMT$) transformations~\cite{Mazzucato:2008tr}, or $TsT$~\cite{Maldacena:2008wh} or the $AdS$ geometry solution of the type $IIB$ supergravity, which has a dual QFT with conformal symmetry, the spacetime can be generated with the Schr\"{o}dinger symmetry which has a dual non-relativistic QFT. Following~\cite{Ammon:2010eq}, we study the DBI action of the probe $D7$ branes in the Schr\"odinger background. From Legendre transformation of the DBI action, we propose the effective action and Euler-Heisenberg effective Lagrangian of the systems with strong interaction with the $z=2$ Schr\"odinger symmetry.
The electric field on the probe branes will distance the two ends of a string on the same D7-brane, and if it is larger than a critical value, it will tear the string apart. In other words, meson dissociates, instability occurs, and non-zero current produces. In the gravity side, this means that the probe branes fall in the background black holes. At non-zero charge density, this always occurs in the AdS background {see\cite{Kobayashi:2006sb}.
	We check this statement in the supergravity Schr\"odinger background.
	On the other hand, the presence of the electric field on the branes introduce world-volume horizon, to which we could assign a temperature. This temperature is different from the background Hawking temperature. Thus, we deal with a non-equilibrium situation. Consequently, the occurrence of instability means that we switch from an equilibrium state to a non-equilibrium one. We study the decay rate of this instability that might produce the fermion and anti-fermion pairs through the Schwinger effect in the Schr\"odinger background. Generally, the instability at the presence of a constant electric field transforms us from an insulator state into a conductor state.
	We study the breakdown of the vacuum (ground state) of strongly coupled systems with the $z=2$ Schr\"{o}dinger symmetry at the presence of an external electric field via probe branes holography.

\section{Review on probe branes in Schr\"odinger background }
Consider a QFT with external conserved current operator $J^a$ which has the $z=2$ Schr\"odinger symmetry.
The $z=2$ Schr\"odinger symmetry respect the Lifshitz scaling as
\begin{equation}
t\to \lambda^2 t \qquad and \qquad \vec{x}\to \lambda \vec{x}.
\end{equation}
The holographic dual to this system could be generated by the null Melvin twist(NMT) transformations of the $D7$ branes as the probe in the background of $D3$ branes. Due to the probe limit, the dynamics of the system is given by the DBI action~\cite{Ammon:2010eq}.  

The DBI action is 
\be\label{DBI0}
S_{D7}\equiv -\Nf T_{D7}\int d\xi^8    e^{-\Phi} \sqrt{{\det}\left(\left[g+B\right]_{ab}+\left(2\pi\alpha'\right)F_{ab}\right)}  \,,
\ee
where $\xi^a$ are $ D7$ worldvolume coordinates and 
\begin{align}
T_{D7}=\dfrac{1}{(2\pi)^7 g_s \a'^4} .
\end{align} \label{Tens}
is $D7$ branes tension.
The  $g_{ab}$ and $B_{ab}$ are the  induced metric and  induced $B$ field from the background on the probe branes, respectively. 
Let embed the $D7$ branes in $10$ dimension space-time as follows :
 \begin{table}[H]
    \centering
    \begin{tabular}{|c|c|c|c|c|c|c|c|c|c|}
        \hline
        &$x^+$&$x^-$&x ,y & r&$\a_1$&$\a_2$&$\a_3$&$\theta$&$\chi$\\
        \hline 
        D3&$\times $& $\times $ &$\times $ & &&&&&\\
        \hline
        D7&$\times $ & $\times $ & $\times $&$\times$&$\times$&$\times $&$\times $&&\\
        \hline
    \end{tabular}
    \caption{\label{tab:1} $D3-D7$ embedding.}
\end{table}
The dual theory would live on intersection of $D3$ and $D7$ branes at $r=0$ which is denoted by the $(x^+,x^-,\vec{x})$. As it clear, there is a $O(2)$ symmetry in $(\chi,\theta)$ direction which clarify the shape of $D7$ branes relative to the background. Without loss of generality, we assume that $\chi =0$ and $\theta=\theta(r)$. We introduce
following gauge field on the probe branes
\be\label{eq:gauge}
A_x=E_b x^+ -2b^2 E_b x^-+a_x(r)\,.
\ee
So the Eq.\eqref{DBI0} would be \footnote{Which we normalized it with volume of boundary theory,i.e.,$Vol_{x^+,x^-,x,y}$}
\begin{align}
\label{DBI1}
S_{D7} = -\mathcal{N} \, \!\!\int\!\!dr \sqrt{-K(r)\,{\det}M_{ab}} \,,
\end{align}
where 
\begin{align}
{\det}M_{ab} \equiv g_{xx}\,g_{\a_1\a_1}\bigg\{
g_{rr}\!\left[(2\pi \a')^2\Eb^2\,H_1 + g_{xx}H_2\right] + (2\pi \a')^2 H_2  a_x'^2
\bigg\} \, ,
\end{align}
which we have defined 
\be \label{hs}
g_{rr}=\frac{1}{r^2 f(r)}+\theta '^2(r)\,,\quad
H_1
= \frac{\b^2[r^2 - \b^2 f(r) \sin^2\!\theta(r)]}{4r^4 K(r)}\cos^4\!\theta(r)\; , \qquad 
H_2 = -\frac{f(r)\cos^4\!\theta(r)}{16r^4 K(r)} \,.
\ee
Before going further let us take a look at  thesolution of Euler-Lagrange equations of $a_x(r)$ and also $\theta(r)$ at the near boundary. Considering zero expectation value for $a_x$ at the dual boundary theory, we have~\cite{Karch:2007pd,Ammon:2010eq}
\be\label{gauge:bdry}
a_x(r)= \frac{\Qx}{2 (2\pi \a')^2 \mathcal{N}}\, r^2+...
\ee
\be\label{theta:bdry}
\theta(r)= 2\pi\a' m r+\theta _2 r^3+...
\ee
where $\Qx$ is conserved current of charged fermions (or flavors in the fundamental representation) in the dual boundary field theory and $m$ is representing the mass of the fermions.
It was shown that at the zero electric fields on the $D7$ branes in the $AdS$ background, there are two allowed embedding for probe branes which could be classified by the ratio of the flavors mass and the background temperature, $(\frac{m}{T})$. The DBI action as the free energy of dual theory would tell us which embedding is thermodynamically favorable \cite{Mateos:2007vn,CasalderreySolana:2011us}. For the large value of $\frac{m}{T}$ the Minkowski embedding (ME) and for small $\frac{m}{T}$ the black hole embedding (BE) is favorable. From a geometry point of view, the ME will happen if the compact dimension of the probe $D7$ branes shrinks to zero outside of the background event horizon. The BE embedding, as its name is, will happen if the compact coordinates fall into the background Blackhole. For the non-zero electric field on the probe branes, we also have another class of embedding~\cite{Mateos:2007vn,Ali-Akbari:2013hba} which known as Minkowski embedding with the horizon (MEH). At the non-zero electric field, the probe $D$-branes would have the world-volume horizon which, in general, differs from the background event horizon. The same embeddings is allow in the Schr\"odinger space-time. For example see Figure~\eqref{fig:1}.
\begin{figure}[h!]
    \makebox[\textwidth][c]{\includegraphics[width=1.\textwidth]{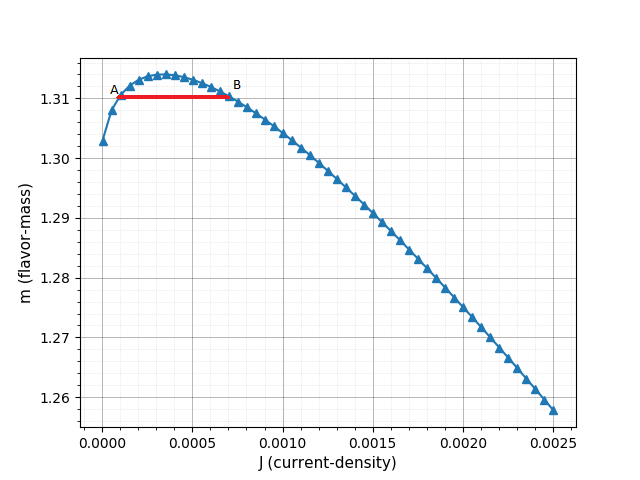}}%
    \caption{$m$-$J$ curve for $\Eb=0.1$ and $T=0.45015 $ and $\b=1$. the maximum value is located at $m_{max}=1.31401$.}
    \label{fig:1}
\end{figure}
It is clear that, from Figure~\eqref{fig:1}, in the small current region we have two current $J$ with the respect to the one $m$~\footnote{For example, see $A$ and $B$ in Figure~\eqref{fig:1}.}. These two different current are illustrating the MEH and BE, for more detail see~\cite{Mobin}. To explain these embedding let us solve Euler-Lagrange equation for $a_x(r)$:
\be\label{eom2}
\frac{\partial \mathcal{L}}{\partial a'_x}=\Qx \to \mathcal{N} K(r)(2\p \a')^2 H_2 \frac{ g_{xx} g_{\a_1\a_1}}{\sqrt{-K(r)\,{\det}M_{ab}}}=\Qx
\ee
Solving $a_x'$ from Eq~ \eqref{eom2} and then inserting it into the Eq~\eqref{DBI1} we get the on-shell action
\be
\label{onsh}
S_{D7}= -\mathcal{N}^2 \int_{0}^{r_H}\!  dr \  K(r)g_{xx}\ g_{\a_1 \a_1}
\ g_{rr}^{\frac{1}{2}} \ \sqrt{\frac{V(r)}{U(r)}}
\ee
which we define
\be
\label{uj1}
U(r) = \frac{\Qx^2}{(2\pi \a')^2\ H_2} + \mathcal{N}^2 K(r)\,g_{xx}\,g_{\a_1\a_1} \, \qquad
V(r)=g_{xx}|H_2| - (2\pi \a')^2\Eb^2\,H_1 \, .
\ee

We could also define the effective action of the dual QFT~\footnote{In the vacuum or equilibrium case this would be free energy, but in here we deal with non-equilibrium steady state,~\cite{Nakamura:2010zd,Nakamura:2012ae,Ali-Akbari:2013hba}.} by the Legendre transformation of the on-shell action~\cite{Hashimoto:2013mua},which is
\be
\label{leg2}
\begin{split}
    \mathcal{L}&=S_{D7} -a_x'  \frac{\delta S_{D7}}{\delta a_x'} \\
    &=-\int_{0}^{r_H}\!\!dr g_{rr}^{1/2}
    \sqrt{V(r)U(r)}\,.
\end{split}
\ee
Obviously from the Eqs.\eqref{uj1}, $U(r)$ and $V(r) $ could be negative or positive. The real condition of the action force  $U$ and $V$ to change their signs at the same point where we call it $r_*$($0<r_*<r_H) $. This point extracts from the following equations:
\begin{gather} 
U(r_*)=0\to \bigg[g_{xx}|H_2| - (2\pi \a')^2\Eb^2\,H_1\bigg]\bigg\vert_{r_*} = 0 ,\,\; \;\notag  \\
V(r_*)=0\to \bigg[ \frac{\Qx^2}{ (2\pi \a')^2 H_2} + \mathcal{N}^2 K(r)\,g_{xx}\,g_{\a_1\a_1}\bigg] \bigg\vert_{r_*}=0 \label{eq:con2}
\end{gather}
The $r_*$ is representing the world-volume horizon or horizon of the open string metric, see Eq.\eqref{eff:metr} and also~\cite{Kundu:13,Kundu1:15,Kim:2011zd}. It could be assign a temperature to this  effective horizon,  which is different from background Hawking temperature  $T$. This situation  shows that we deal with a non-equilibrium condition. In other words, the matter sector, which realize by the probe $D7$ branes, has different temperature relative to the background plasma, $D3$ branes. Therefore, in the dual theory we deal with non-equilibrium steady state,~\cite{Nakamura:2010zd,Nakamura:2012ae}. \\
Continue with the Eq.~\eqref{eq:con2}, we able to derive the nonlinear $DC$ conductivity, $\Qx=\sigma(\Eb) \Eb$ :
\be \label{conduc}
\sigma =\frac{\mathcal{N}\, b \cos^3\theta(r_*)}{4 r_*^2 }\sqrt{r^{2}_*-b^2 \sin^2\theta(r_*) f(r_*)}\, \,  \quad .
\ee
\\
Interestingly, as it clear from the Figure~\eqref{fig:1}, for the mass of fermions greater than a maximum value, $m_{max}$, the current $\Qx$ is zero, and we have Minkowski embedding. The phase transition could occur, and the state with $\Qx=0$ switches to $\Qx\ne0$  state or from ME to BE (or MEH), see\cite{CasalderreySolana:2011us,Mateos:2007vn} . 
\\
Let us forget about non-equilibrium condition and consider $\mathcal{L}$  in the Eq~\eqref{leg2} as the Helmholtz free energy same as the equilibrium thermodynamic. The heat capacity i.e.,
\be\label{hc}
C_V=-T \frac{\partial^2  \mathcal{L}}{\partial T^2}\bigg \vert _{\Eb,\b}
\ee
feels singularity exactly at the $r_*$ in Eqs.~\eqref{eq:con2}, likes to the first order phase transitions in ordinary thermodynamics.\\
For the zero electric fields, we do not have any world-volume horizon or the singularity for heat capacity~\eqref{hc}. One can conclude that: At non zero temperature, the electric field breaks the bond state of neutral charge pairs, which are binding as mesons or Cooper pairs. Also at the zero background temperature, this phenomenon happens because of the external electric field and existence of MEH. A transition from the state with$\textcolor{\tcolor} {\Qx =0}$ in the dual theory to the states with $\textcolor{\tcolor}{\Qx \neq0}$ can be considered as a change from false or metastable ground state (or vacuum) to a true ground state. Consequently, at the presence of the electric field an insulator state $\Qx =0$ will suffer from the instability due to the electrical breakdown. Although we will see that, on the Schr\"odinger background, there is a situation to have a perfect insulator.
%
In the next section we investigate the instability which causes the phase transitions in schr\"odinger geometry from type IIB supergravity.
 \section{Ground state instability}
 From the holographic point of view in the AdS background, the imaginary part of an effective action, which shows vacuum to vacuum transition, has been studied in \cite{Hashimoto:2013mua} for the supersymmetric QCD (QED) systems. It would be interesting if we could generalize this idea to other systems such as condensed matter systems, by using gauge/gravity duality. 
Following  \cite{Hashimoto:2013mua}, for a system with Schr\"odinger symmetry such as $cold \, atoms$, we study the decay rate of the ground state to ground state via probe branes holography.
To do this we put $\Qx=0$ in the Eq.\eqref{leg2} or Eq.\eqref{onsh}\footnote{For $\Qx=0$ , Eq.\eqref{leg2} is a same as Eq.\eqref{onsh}}, which means that the electric field is turned on and we have our probe branes with Minkowski embedding yet. In other words we are studying  electric field effects on the insulator state which is a pair of fermions bound together. In the condensed matter environment  the dielectric breakdown of band(or Mott) insulator from the ground state to ground state transition was studied in~\cite{oka:514}.  As repeatedly mentioned, living in an insulator state means we have $\Qx=0$ so from Eq.\eqref{leg2} we will have:
\be\label{act:zeroJ}
\mathcal{L} =-\mathcal{N} \int_{0}^{r_H}\!\! dr K^{1/2}(r)\sqrt{g_{xx}g_{rr}g_{\a_1\a_1}\left[g_{xx}|H_2| - \pa^2 \Eb^2\,H_1\right]}  .
\ee
It is clear that the function under the square root in the Eq.\eqref{act:zeroJ} can be a negative quantity at specific intervals.
Therefore,  the effective action is a complex quantity and
  in general we have
\be \label{eff:n}
\mathcal{L} = i \, \Im\mathcal{L}  \bigg \vert_{r_I}^{r_H}+\Re \mathcal{L} \bigg \vert_{0}^{r_I}
\ee
The $r_I$ ($0<r_I<r_H$) is obtained from the following equation
\be\label{eq:imagr}
\left[g_{xx}|H_2| -(2\pi \a')^2 \Eb^2\,H_1\right]_{r_I}=0\,.
\ee
Definitely, at the zero electric fields, we will have $r_I=r_H$.
Complex effective action is a symbol of having an instability of the system which means that the system lives in the false vacuum or false ground state. In a Quantum theory, the vacuum to vacuum amplitude is provided by
\be \label{ampli}
\langle 0|U(t)|0\rangle \propto\exp(i \, \mathcal{L}\; V t) \nn
\ee
where $U(t)$ is unitary time evolution operator of the system and $V $ is a volume of the space and $|0\rangle$ stands for a ground state.In general we have 
 \be \label{eff0}
 \mathcal{L}=\mbox{Re} \mathcal{L}+i \frac{\Gamma }{2}\,.
 \ee
 Therefore, the non-zero imaginary part of effective action, same as Eq.\eqref{eff:n}, is proportional to the amplitude of decay rate of the vacuum. As previously discussed, at the presence of an external electric field  decay of the unstable vacuum to the stable vacuum can be interpreted as the \textit{Schwinger-like} pair production in the Schr\"odinger geometry\footnote{Instead of \textit{vacuum} we repeatedly use\textit{ ground state} due to the non-zero chemical potential $\mu$ which in the dual theory related to the number operator of a Schr\"odinger algebra.}. In the following sections, we study the imaginary and the real part of the effective action for the massless and massive charge carriers in the Schr\"odinger background and compare our result with the relativistic one in the AdS background.
\section{ Ground state instability for gapless systems }
For the embedding with $\theta(r)=0$ in the bulk gravity side we would have massless charge carriers in the dual boundary theory. Since at the zero mass we deal with the scale invariant theory\footnote{At the zero temperature.}, this configuration resembles the gapless systems in the condensed matter systems\footnote{In condensed matter physics it was suggested that Kondo insulators are gapless, see~\cite{gapless}.}. For the $\theta(r)=0$ Eqs.\eqref{hs} reduce to
\be \label{hs0}
H_1
= \frac{\b^2 }{4r^2 K(r)} \, , \qquad 
H_2 = -\frac{f(r)}{16r^4 K(r)} \,.
\ee
Hence, from the Eq.\eqref{eq:imagr} we find that  
\be\label{masslessri}
r_{I}= \kappa\,r_{H} 
\ee 
which we  have defined that
\be\label{kappa}\begin{split}
    \kappa=\Big(1+\,4 \pa^2 \Eb^2 \b^2 r_{H}^4 \Big)^{-1/4}
    =\big(1+ 16\Eb^2 \dfrac{|\mu|}{\pi^2 T^4 \lambda }\big)^{-1/4}  .
\end{split}
\ee
In the last term we apply Eq.\eqref{tmu} and $\pa^2=2\pi^2\lambda^{-1}$. Clearly $\kappa<1$ so as already mentioned we have complex effective action with: $0< r_I<r_H$.
If we restore the $\Eb$ with $E/2\b$, Eq.\eqref{masslessri} will coincide with $r_I$ for massless flavors in the $AdS$ background, see~\cite{Hashimoto:2013mua}.
\subsection{Decay rate of ground state for the gapless systems}
As previously discussed, the imaginary part of an effective action is related to the decay rate of the systems from false vacuum to the true vacuum. From Eq.\eqref{eff:n} and Eq.\eqref{act:zeroJ} for the massless bonded fermions we get that
\be 
\begin{split}
    \Im    \mathcal{L} &=-\mathcal{N} \int_{r_I}^{r_H} \frac{dr}{8\ r^5}\sqrt{\frac{-f(r)+(2\pi \a')^2 4 \b^2\Eb^2 \ r^4}{f(r)}}\\ &
    =-\frac{\mathcal{N}}{8 \kappa^2 r_{H}^4}(\frac{1-\kappa^4}{8\kappa^2} \pi)\,.
\end{split}
\ee 
After replacing $\kappa$ from Eq.\eqref{kappa}, we will find that
\be
\Im\mathcal{L}= \frac{\mathcal{N} \pi}{64}(2\pi \a')^2 4 \b^2 \Eb^2 \quad .
\ee
Remembering that  $\mathcal{N}=N_{f} T_{D7} 4\pi^2$ and also
\be
\lambda=N_c\, g_{QFT}^2\qquad
 2\pi g_s=g_{QFT}^2  \qquad T_{D7}=\frac{1}{(2\pi)^7 g_s \a'^4},
\ee
from Eq.\eqref{eff0},we get that
\be\label{im0}
\Gamma=\frac{N_f N_c}{32 \pi }4\b^2\Eb^2\,=\frac{N_f N_c}{16 \pi |\mu| }\Eb^2\,.
\ee
This is similar to the pair production amplitude from Schwinger  instability in QED, see~\cite{Schwinger:1951nm}. The Eq.\eqref{im0} is the same as $AdS$ result  in~\cite{Hashimoto:2013mua},
 if we replace $\Eb=E/2\b$ and $\Nf=1$.
 The reason for the similarity result between relativistic and non-relativistic Schwinger instability is that they have the same bulk mechanism of instability; the electric field will tear apart strings with both ends on the same probe brane. So finally we could say that for the $\Eb\ne0$ the system always will decay from $\Qx=0$, an insulator state, to $\Qx\ne0$ or a conductor state. This result, as Eq.\eqref{im0} shows it, is independent of the background temperature. Therefore, at zero background temperature, we also will have the ground state instability, and the Minkowski embedding will switch to the other embedding with non-zero current which is MEH. Moreover, it is evident from Eq.\eqref{im0} that\textit{ it does not matter how small the electric field is}, the ground state of massless fermions is always unstable because of the electric field.At the next section, we will see that the ground state in the electric field for the massive fermions behaves entirely different to the one for massless fermions.
\subsection{Euler-Heisenberg action for the gapless systems}
It was shown that the real part of the effective action will produce the Euler-Heisenberg effective action of the QED(SQCD),~\cite{Hashimoto:2013mua}. In the Schr\"odinger background from Eq.\eqref{act:zeroJ}, the real part of the effective action for massless embedding would be 
\begin{align}
\Re \mathcal{L}=-\frac{\mathcal{N}}{8 \kappa^2 r_{H}^4} \int_{0}^{\kappa}  dx \frac{1}{x^5} \sqrt{\frac{\kappa^4-x^4}{1-x ^4}}
= -\frac{\mathcal{N}}{8 \kappa^2 r_{H}^4}\bigg(\frac{\sqrt{\pi } \kappa^3 \Gamma \left(\frac{5}{4}\right) \, _2F_1\left(\frac{1}{4},\frac{1}{2};\frac{7}{4};\kappa^4\right)}{2 \Gamma \left(\frac{7}{4}\right)}\bigg).\label{real:t}
\end{align}
 For small $\kappa$ or $\mu \, \Eb^2 <  \lambda\, T^4$ we have
\be \label{real:tt}
\Re \mathcal{L}=\mathcal{N}\big( c_{1} + c_{2}|\mu| \, \Eb^{2} \, \ln (\frac{\Eb^{2}|\mu|}{\lambda T^4}) + \mathcal{O}((\frac{\Eb^2 |\mu|}{\lambda T^4})^2) \big)
\ee
where $c_1$ and $c_2$ are numeric constant. Again if we redefine $E=2\b \Eb$ the similar result to the relativistic gapless systems will produce, see \cite{Hashimoto:2013mua} and references therein. The real part of the effective action depends on the background  temperature. We could find a difference between the real effective action in the conducting phase $\Qx\ne0$, Eq.\eqref{leg2}, and in the insulator phase, Eq.\eqref{real:tt}, which would be a positive and finite quantity: 
\be\label{criteria}
\Re \mathcal{L}_{insulator}- \Re \mathcal{L}_{Conductor}\propto  C \b^2\Eb^2
\ee
where $C$ is a positive numeric constant. Thus we could presume that the energy difference is also finite so the real part of the effective action for zero current is meaningful, see also~\cite{Hashimoto:2013mua}.

\section{Ground state instability for the gapped systems}
By the gapped systems we mean  that we would deal with massive charge carriers. For simplicity we consider massive fermionic degrees of freedom at the background medium(plasma) with zero temperature. As already discussed, for massive flavors we have $\theta(r)\ne0$\footnote{We consider the distance between D3 and D7 branes such that $\sin\theta(r)=2\pi \a' m$}. With this assumptions the  induced metric on the probe $D7$ branes, with Table.\eqref{tab:1} embedding, is 
\be \label{mass:metric}
ds^2= \frac{M(r)}{r^2}\Big(-\frac{M(r)dx^{+2}}{r^2}+2 dx^+ dx^- +dx^2+dy^2 \Big) +\frac{1}{M(r)} \left(dr^2 /r^2+ds_{\alpha}^2\right)
\ee
where $M(r)=1+(2\pi \a' m)^2 r^2$ and $ds_{\a}^2=(\sigma_{1}^2+\sigma_{2}^2+\sigma_{3}^2)$\footnote{This is quit similar to $ds_{S^3}^2$ . }.
For $m=0$, the metric Eq.\eqref{mass:metric} would have $z=2$ Schr\"odinger isometry.
With a little bit of work we recover the Eq.\eqref{act:zeroJ} with 
\be
H_2=\frac{-1}{16r^4}  \,\, ,\qquad H_1=\frac{\b^2(1-( 2\pi \a' m)^2 \b^2)}{4r^2 M(r) } \qquad.
\ee
The instability condition i.e.,Eq.\eqref{eq:imagr}, simplifies to
\be \label{mass:imagr}
1-\frac{4(2\pi \a')^2 \b^2\Eb^2\Big (1-( 2\pi \a' m)^2 \b^{2}\Big) r_I^4}{M^2(r_I)}=0\,.
\ee 
\paragraph{Note:}For $( 2\pi \a' m)^2 \b^2<1 $ we could always define a real effective electric field $\tilde{E}$ such that $\tilde{E}^2=4\b^2\Eb^2 (1-( 2\pi \a' m)^2 \b^{2})$; therefore, we could  rewrite the Eq.\eqref{mass:imagr} as  
\be
1-\frac{(2\pi \a')^2\tilde{E}^2}{M^2(r_I)}=0\,.
\ee
This equation will give us $r_I,$ which looks similar to the result in the $AdS$ background for QCD like systems:
\be \label{absm}
r_{I} ^2=\sqrt{2\pi \a'\tilde{E}-(2\pi \a' m)^2} \, .
\ee
The new electric field $\tilde{E}$ has dimension of the relativistic electric field. Clearly we would have real $r_I  $ if the electric field has a larger value than the critical value which is $ \tilde{E}_c= 2\pi \a' m{}^2 $~\footnote{We should note that $E_c =2\pi \a' m^2$ is the critical electric field in the AdS background, see~\cite{Hashimoto:2013mua}.}. So the critical electric field $E_c$  is
\be \label{crit:rel}
E_c=\frac{2\pi \a' m^2}{\sqrt{1-\b^2 (2\pi \a' m)^2}} \,. 
\ee
or we could define a non relativistic critical electric field $\Eb ^c $ as follows
\be \label{crit}
\Eb^c=\frac{2\pi \a' m^2}{2\b\sqrt{1-\b^2 (2\pi \a' m)^2}} \,. 
\ee
If we compare the critical value of electric field in Eq.\eqref{crit:rel} with the relativistic one which is $E_c=2\pi\a' m^2$~\cite{Hashimoto:2013mua}, we see that 
\footnote{Note that we considered $( 2 \pi \a' m )^2 \b^2 <1$ or in term of $\mu$ and $\lambda$, $\frac{m^2}{\lambda}<\frac{|\mu|}{\pi^2}$.}
\be \label{bud}
E_c ^{sch}> E_c ^{AdS},
\ee
which means that in the non-relativistic systems the external electric field must be stronger than its relativistic counterpart to pairs production happen or instability occurs\footnote{We must notice that electric field in the relativistic theory has a different scale dimension in comparison to theory with Schr\"{o}dinger symmetry, so we compare  $ 2\b\Eb^c= E_c^{sch}$ and $E_c^{AdS}$.}.
By defining the \textit{effective mass} $m_{*}$ as :
\begin{align}\label{eff:mas}
m_{*}^2=\frac{m^2}{\sqrt{1-\b^2 (2\pi \a' m)^2}} \,,
=\frac{m^2}{\sqrt{1-\frac{\pi^2  m^2}{\mu \lambda}}}
\end{align}
from the Eq.\eqref{crit:rel}(or Eq.\eqref{crit}), we will have
\be\label{crit:sch}
E_{c}^{Sch}=2 \b \Eb^c= 2 \pi\a' m_{*}^2  .
\ee
This is the same as the relativistic critical electric field which the mass of fermion $m$ has been replaced by the effective mass $m_{*}$. Consequently, the geometric distinction between AdS and Schr\"{o}dinger space-time  has been changed to the difference in masses of  charge carriers in the dual theory. Therefore, we could propose that the potential between two fermions or quarks with a distance of $l$ from each other, in the Schr\"odinger space-time Eq.\eqref{eq:schmetric}, would be:
\begin{align}
V(l)=\frac{\sqrt{2} \pi}{\sqrt{\lambda}} \frac{m_{*}^2 }{2\b \, l}\,.
\end{align}
This result is nontrivial, but due to the compact coordinate $x^-$, would make sense\footnote{It would be interesting to find this potential from the Wilson loop calculation, but for finite quark mass there exist difficulties. For the fermions with a large mass see~\cite{Alishahiha:2003ru,Araujo:2015dba,Akhavan:2008ep}.   }. In the solid-state physics, moving an electron inside a crystal lattice would be the same as its motion in the vacuum if we use effective mass for the electron instead of the electron's mass; the similar behavior is observed in here. 
Due to the effective mass, we could assume that the compactification along the $x^-$ brings the same physics as a periodic potential brings to the study of the band structure in a crystal lattice at solid-state physics. It should be clear that from Eq.\eqref{eff:mas}, this is significant effect at the finite 't Hooft coupling $\lambda$ and for large 't Hooft coupling, $m_*= m$; therefore, we are reduced to the result in the AdS background.

\subsection{Decay rate: Imaginary part of the effective action}
The effective action for the gapped system or massive fermions, from Eq.\eqref{act:zeroJ} and Eq.\eqref{eff:n}, for the electric field higher than the critical electric field, i.e.,$\Eb>\Eb^cc$, has imaginary term which is given by 
\begin{align}\label{eff:mass}
\Im \mathcal{L} =- \mathcal{N} \int _{r_I}^\infty{} dr  \frac{1}{8 r^5}\sqrt{-1+\frac{(2 \pi \a')^2\tilde{E}^2 r^4}{M(r)^2}}
\end{align}
This is the quite same effective action that was found in the $AdS$ background in which $E$ is replaced by $\tilde{E}$.
With subsequent changes
\be
\frac{r^2}{r_{I}^2}=1+x  \qquad , \qquad \epsilon =\frac{\tilde{E}_c}{\tilde{E}}=\frac{2\pi \a' m_*^2}{2\b\Eb} \,,\nn
\ee  
Eq.\eqref{eff:mass} changes to 
\be
\Im\mathcal{L}=\mathcal{N}\frac{(1-\epsilon)^{5/2}(2\pi \a')^2 \tilde{E}^2}{16 } \int_{0}^{\infty} dx \frac{\sqrt{x(2+x+\epsilon x)}}{(1+x)^3 (1+\epsilon x)}
\ee
for large electric field respect to $\tilde{E_c}$ we could use the $\epsilon$ expansion which would give us following result 
%
\be
\Im \mathcal{L}= \mathcal{N}\frac{\pi}{64} (2\pi \a')^2 \tilde{E}^2\Big(1+\frac{\pi}{4}\frac{\tilde{E}_c}{\tilde{E}} Log \frac{\tilde{E}_c}{2 \tilde{E}} -\frac{1}{3\pi} (\frac{\tilde{E}_c}{2 \tilde{E}})^3+\mathcal{O}((\frac{\tilde{E}_c}{2 \tilde{E}})^ 4)\Big)\,,
\ee

This quantity is the same as the relativistic result from $AdS$ background, see~\cite{Hashimoto:2013mua}.
Expanding imaginiary term of the effective action relative to the small $\frac{\pi^2 m^2}{|\mu| \lambda}$, and also replacing $E=2\b \Eb$, the decay rate or pair production probability will be
\begin{equation} \label{im:m}
\Gamma_{Sch}
=-\pi \frac{\Nf \Nc m^2}{ \lambda |\mu|} E^2\big(1-\frac{m^2}{8 \sqrt{\lambda}E}\big) +\big(1 + \frac{m^2 \pi^2 }{\lambda |\mu| }\big) \Gamma_{AdS}+\dots
\end{equation} 
where 
\be\label{im:ads}
\Gamma_{AdS}= \mathcal{N}\frac{\pi}{32} (2\pi \a')^2 E^2\Big(1+\frac{\pi}{4}\frac{\tilde{E}_c}{E} Log \frac{\tilde{E}_c}{2 E} -\frac{1}{3\pi} (\frac{\tilde{E}_c}{2 E})^3+\mathcal{O}((\frac{\tilde{E}_c}{2 E})^ 4)\Big)\,.
\ee
It is explicit that for the zero mass, or infinite 't Hooft coupling the Eq.\eqref{im:m} will reduce to the decay rate in the zero temperature AdS background, i.e., Eq.\eqref{im:ads}.
For the massless particles in the zero temperature AdS spacetime, we know that the $DBI$ action for the probe $D7$ branes does not change under NMT transformations, see~\cite{Ammon:2010eq}, but 
  the $DBI$ action for the massive particles will change under the NMT transformations, so the result in Eq.\eqref{im:m} make sense. It is clear that if the particle's mass goes to zero, we will get back to $AdS$ result, due to the same effective action. 
    The dependence of the imaginary part of the effective action to 't Hooft coupling also is shown in~\cite{Taghavi} for chiral mesons at the presence of external electromagnetic fields, see also~\cite{bitaghsir, Ghodrati}. Other terms in Eq.\eqref{im:m}, which differ from $AdS$, might be considered as a dipole interaction that inherently exists in the dual theory of this Schr\"odinger spacetime which is originated from type IIB supergravity, see~\cite{Alishahiha:2003ru} and references therein.
\subsection{Euler-Heisenberg action: real part of the effective action for gaped systems }
As already mentioned the real part of the effective action is related to the Euler-Heisenberg action. For the massive particles at zero temperature from Eq.\eqref{eff:n} we will have
\begin{align} \label{effr:mass}
\Re \mathcal{L}=- \mathcal{N} \int _{0}^{r_I} dr  \frac{1}{8 r^5}\sqrt{1-\frac{(2 \pi \a')^2\tilde{E}^2 r^4}{M(r)^2}}=\mathcal{N}\frac{(1-\epsilon)^{5/2}(2\pi \a')^2 \tilde{E}^2}{16 } \int_{0}^{1} dy \frac{\sqrt{y(2-y-\epsilon y)}}{(1-y)^3 (1-\epsilon y)}\,
\end{align}
where we have defined $y=1-\frac{r^2}{r_{I}^2} $.
Again, this is similar to real part of effective action in $AdS$ background \cite{Hashimoto:2013mua}. For small $\epsilon$ i.e., Strong electric field $\tilde{E}$ relative to $\tilde{E_c}$, the critical electric field, the finite part\footnote{After regularization.} of the \eqref{effr:mass} would be
\be \label{fin:eff}
\Re \mathcal{L}= \frac{\Nc \Nf}{32 \p^2} \, \tilde{E}^2\bigg(3+\ln 2\, +\ln \frac{\tilde{E}^2_{c}}{\tilde{E}^2}+\frac{\tilde{E}^{2}_{c}}{\tilde{E}^2}+\frac{\p \tilde{E_c}}{E} \bigg).
\ee
Clearly if we insert $\tilde{E}=\frac{\Eb}{\sqrt{2 |\mu|}}  \sqrt{1-\frac{\pi^2 m^2}{|\mu| \lambda}}$ and $\tilde{E}_c=2\pi \a' m^2$ in Eq.\eqref{fin:eff},or $\frac{\tilde{E}_c}{\tilde{E}}=\frac{2\pi \a' m_*^2}{E}$, there would be correction terms, compared to the outcome in the $AdS$ spacetime, in term of   $\frac{\pi^2 m^2}{|\mu| \lambda}$ which are relevant at the finite 't Hooft coupling. However, at the zero mass, we will have the same Euler-Heisenberg effective Lagrangian which is found in the AdS background.
At zero temperature for the massive neutral charge carriers, there are also two different situations that the effective action is not a complex quantity, and always remains real at the presence of an external electric field. In other words, the Minkowski embedding $\Qx=0$ is a stable solution, and the electric field is not strong enough to break the bond between neutral charge pairs. One of them will exist if we have an electric field below the critical electric field, i.e., $\tilde{E}\leq\tilde{E}_c$. This real effective action lives in both AdS background and Schr\"odinger background.
\paragraph{Stable bound state:}From the Eq.\eqref{mass:imagr}, we will have real effective action regardless of the electric field if we suppose that
\be \label{conr}
1-(2\pi \a'\,m)^2\b^2<1 \quad or\quad \frac{m^2}{\lambda}>\frac{|\mu|}{\pi^2}\quad .
\ee
Considering Eq.\eqref{conr}, the effective mass in Eq.\eqref{eff:mas} will be an imaginary quantity. This does not have a relativistic or $AdS$ counterpart. Interestingly this condition does not depend on the electric field; therefore\textit{ it does not matter how large the electric field is}, at the regime of Eq.\eqref{conr}, the effective action is always a real quantity. If we replace 't Hooft coupling $\lambda$ with $e^2$, to extract QED-like results following~\cite{Hashimoto:2013mua}, one could say that for the neutral charge pairs with the mass-to-charge $\frac{m_e}{e}$ ratio larger than $\frac{1}{\sqrt{2}\pi\, \b }$ we have \textit{permanent or perfect insulator}. In this case, the effective action is an addition of the two real quantity, Eq.\eqref{eff:mass} and Eq.\eqref{effr:mass}. If one does not accept the existence of perfect insulator at any circumstances, one can put the upper limit for the fermion's mass, and say that the insulator ground state will decay to the conductor ground state. For the massive particles at non-zero temperature, due to the analytic difficulties, we need numerical calculations, but naively from Eq.\eqref{conduc} we could argue that the conductivity  might always be zero, so current $\Qx$ could be zero if 
\begin{equation}
r_*^2-b^2 f(r_*) \sin^2\theta(r_*)=0.
\end{equation} 
Nevertheless, the $r_*$ and $\theta(r_*)$  depend on the electric field and mass of the fermions, see Eqs.\eqref{eq:con2} and see also~\cite{Mobin}. Hence, we are able to find a critical electric field which depends on the mass of the particles and background temperature. Therefore it can be concluded that at a non-zero temperature, a perfect insulator will not exist according to this argument, and the insulator state will decay to the conductor state.
\section{Conclusion and Summary}
We study the breakdown of the vacuum (ground state) of strongly coupled systems with the $z=2$ Schr\"{o}dinger symmetry at the presence of an external electric field via probe branes holography.
By using the holographic argument, the decay rate of the ground state to the stable ground states, which causes the Schwinger pair production, is calculated in a finite 't Hooft coupling. From the gravity side of the duality, there are three embedding classes: Minkowski embedding (ME) which exists if the probe brane closes off the background event horizon, black hole embedding (BE) which exists if the probe branes fall into the background black hole, and Minkowski embedding with the horizon (MEH) which exists at the presence of the non-zero electric field. For both BE and MEH, we have the non-zero current $\Qx\ne0$, and thus the system in the boundary quantum theory lives in the conductor state. For ME, the current is zero, and at the boundary theory, we have an insulator state. We might consider the string with both ends at the same brane as mesons, Cooper pairs or a bonded electron-hole. In this study, we investigate instability at the presence of the constant external electric field when we have $\Qx=0$ or the insulator state in the Schr\"odinger background. For the massless particles or gapless systems in the Schr\"odinger spacetime, both the real and imaginary parts of the effective action look similar to the effective action in the AdS background. For the massive fermions, the decay rate from the insulator to the conductor would be the same as the AdS results if we recall the effective mass $m_*$ instead of the mass $m$ and electric field with $E (1-\frac{\pi^2 m^2}{|\mu| \lambda})^{1/2}$. For an electric field greater than the critical electric field, which is proportional to the square of the effective mass, the ground state or insulator will decay to other ground states or conductors. We show that the false vacuum would be faded out if there is an upper bound for the mass of the massive particles. In other words, the bond between fermions-anti-fermions or quarks-anti-quarks will not break by the electric field if we have particles with the mass larger than $\frac{\sqrt{\lambda\, |\mu|}}{\pi}$. Therefore, the system lives in the insulator phase forever.
The effective action in here looks similar to the relativistic one \cite{Hashimoto:2013mua}, if the electric field is replaced by $E (1-\frac{\pi^2 m^2}{|\mu| \lambda})^{1/2}$.

\section*{Acknowledgments}
We would like to thank S.\;F.\;Taghavi ,  K.\;Bitaghsir and R.\;Mohammadi  and J.\;Khodagholizade and M.\;B\;Fathi  for useful discussions. Special thank to  D.\;Allahbakhshi for reading the manuscript and fruitful comments.

\appendix

\renewcommand{\theequation}{\Alph{section}.\arabic{equation}}
  \setcounter{equation}{0}  

\section{Schr\"odinger spacetime}
The near horizon limit of non-extremal $D3$ branes solution in type $IIB$ superstring theory which is  $AdS_5$ Schwarzschild times $ S^5 $ (with AdS radius  $L$) is
\begin{equation}
    \label{D3}
    ds^2 = \frac{L^2}{r^2} \left( \frac{dr^2}{f(r)} -f(r) dt^2 + dy^2 + d\vec{x}^2 \right) + \L ds^2_{S^5},
\end{equation}
Here $f(r)=1-\frac{r^4}{r_{H}^4}$ which tell us that black hole's horizon located at $r_H$.
The radial coordinate is $r$, and boundary located at $r=0$ and field theory lives on $(t,y,\vec{x})$ which $y$ is singled out because we need to do  null Melvin twist(NMT) operation along with it. The metric \eqref{D3} has  $ISO(1,3)\times SO(6)$  isometry at extremal limit. The holographic dual of this geometry \footnote{there is also five form $RR$ field. } is $\mathcal N= 4$ superconformal $SU(\Nc)$ gauge theory in the large $\Nc$ limit and large 't Hooft coupling($\lambda=\Nc  g_{YM}^2$)\footnote{at zero temperature or extremal limit we have supersymmetry. At non zero temperature, we have a thermal state in a dual field theory where supersymmetry is broken.}. We could write the $S^5$ metric as a Hopf fibration over $\cp$, with $\chi$ the Hopf fiber direction
\be \label{cpp} 
ds^2_{S^5}=\left(d\chi + \mathcal{A} \right)^2 + ds^2_{\cp}
\ee
where  $\mathcal{A}$ gives the K\"ahler form $J$ of $\cp$ via $d\mathcal{A}=2J$.
To write the \eqref{cpp} explicitly, we introduce $\cp$ coordinates $\a_1$, $\a_2$, $\a_3$, and $\theta$ and define the $SU(2)$ left-invariant forms
\bea
\label{eq:cp2}
\s_1 & = &  \frac{1}{2} \left( \cos \a_2 \, d\a_1 + \sin\a_1 \, \sin\a_2 \, d\a_3 \right), \nn \\ \s_2 & = & \frac{1}{2} \left( \sin \a_2 \, d\a_1 - \sin\a_1 \, \cos\a_2 \, d\a_3 \right), \nn \\ \s_3 & = & \frac{1}{2} \left( d\a_2 + \cos\a_1 \, d\a_3\right),
\eea
so we could write the metric of $\cp$ as follows
\be
ds^2_{\cp} = d\theta^2 + \cos^2\theta \left( \s_1^2 + \s_2^2 + \sin^2\theta \, \s_3^2 \right),
\ee
and $\mathcal{A} = \cos^2 \theta \, \s_3$. The full solution also includes a nontrivial five-form, but it's shown in refs. \cite{Herzog:2008wg,Maldacena:2008wh,Adams:2008wt} that five-form will be unaffected by the NMT or TsT.
\\
After the null Melvin twist operation \cite{Alishahiha:2003ru}, we get the following metric\footnote{The $TsT$ transformation\cite{Mazzucato:2008tr} also gives us similar result but a little bit different. In\cite{Alishahiha:2003ru} although the SUSY is broken but 8 supercharges have been remained while in TsT \cite{Mazzucato:2008tr} whole SUSY is broken.}
\bea
\label{eq:schmetric}
ds^2 &=&
 \frac{\L}{r^2} \left( \frac{dr^2}{f(r)} - \frac{f(r)}{\L r^2 K(r)} dx^{+ 2} + \frac{2}{K(r)} dx^+ dx^- + \frac{1-f(r)}{2 K(r)} \left( \frac{dx^+}{\sqrt{2} \b L} - \sqrt{2} \b L dx^- \right)^2 + d\vec{x}^2 \right) \nn \\ & &  \qquad \qquad \qquad \qquad \qquad \qquad \qquad \qquad + \frac{\L}{K(r)} \left(d\chi + \mathcal{A} \right)^2 + \L ds^2_{\cp},
\eea
where
\be
f(r) = 1 - \frac{r^4}{r_H^4}, \qquad K(r) = 1 + \frac{\b^2 r^2}{r_H^4},
\ee
The solution also includes the  Kalb-Ramond
two-form $B$ field
\bea
\label{eq:bfield}
B= 
- \frac{\L}{2 r^2 K(r)} \left(d\chi + \mathcal{A} \right) \wedge \left( \left( 1 + f(r) \right) \,\frac{ dx^+}{L} + \left( 1 - f(r) \right) 2 \b^2 L dx^- \right)
\eea
and also a dilaton
\be
\label{eq:schdilaton}
\Phi = - \frac{1}{2} \log K(r).
\ee
The zero temperature metric will be produced from the  Eq.\eqref{eq:schmetric} by $r_H \rightarrow \infty $, which is
\bea \label{eq=zeroschmetric}
ds^2  = \frac{1}{r^2} \left( dr^2 - \frac{1}{r^2 } dx^{+ 2} + 2 dx^+ dx^- + d\vec{x}^2 \right)    + \left(d\chi + \mathcal{A} \right)^2 + ds^2_{\cp},
\eea
Where we set  $L=1$, just for simplicity. As you see there is no $\b$ in this metric. There is also $B$ field with no dependence on $\b$.
If we do a compactification on $S^5$\footnote{$ds_{S^5}^2=\left(d\chi + \mathcal{A} \right)^2 + ds^2_{\cp}$} the \eqref{eq=zeroschmetric} will be a Schr\"odinger metric which is introduced in \cite{Nishida:2007pj,Son:2008ye,Balasubramanian:2008dm} for gravity dual part of  non-relativistic CFT. In the\cite{Son:2008ye} was shown that the Schr\"odinger geometry
\be \label{met:fremi}
ds^2  = \frac{1}{r^2} \left( dr^2 - \frac{1}{r^2 } dx^{+ 2} + 2 dx^+ dx^- + d\vec{x}^2 \right) \,
\ee
in which $x^+$ is a time coordinate in the dual field theory and for compact $x^-$, could be a dual to the free fermions or fermions at unitarity and in\cite{Balasubramanian:2008dm} was discussed the cold atom aspects. The \eqref{eq=zeroschmetric} or \eqref{met:fremi} will be preserved with the following scale transformations
\be \label{sacle:schro}
x^+  \rightarrow \lambda ^2 x^+ \qquad \qquad  
x^- \rightarrow  \lambda ^0 x^-  \qquad \qquad 
r \rightarrow \lambda  r  \qquad 
\vec{x} \rightarrow \lambda  \vec{x}
\ee
$\b$ is a dimensionful parameter which  has units of length and the $x^+$ has the dimension of square of length i.e., $[L]^2$ and $x^-$ has no dimensions. In  the Schr\"odinger space-time the $x^-$ is a compact dimension so at the boundary $r=0$ we have $2+1$ dimension theory. The isometry generator along $x^-$ is a dual to $numeber$ operator $N$ in dual theory. At finite temperature i.e., \eqref{eq:schmetric}, there would be a momentum along $x^- $($P_-$). So the quantum state in the dual theory has finite number density $N$ or chemical potential\cite{Kim:2010tf,Herzog:2008wg,Adams:2009dm}. As mentioned in\cite{Kim:2010tf,Herzog:2008wg,Adams:2009dm} the temperature and the chemical potential of the dual quantum field theory, which is due to $U(1)$ symmetry along $x^-$ compact direction and not charge carriers, would be
\be \label{tmu}
T=\frac{1}{\pi  r_H  \b L} \qquad \qquad \mu = - \frac{1}{2\b^2 \L}.
\ee
One of the fascinating feature of zero temperature schr\"odinger space-time is that a flavor quark would feel a drag force~\cite{Akhavan:2008ep} and also one of  the interesting comment about a Schr\"odinger metric~\eqref{met:fremi} is that this geometry has a $SL(2,R)$ asymptotic symmetry~\cite{Alishahiha:2009nm}.
\section{Effevtive metric}
DBI action integrand is given by
\be
\det(g_{ab}+A_{ab})
\ee
where we define $A_{ab}=B_{ab}+2\pi \a' F_{ab}$. The $A_{ab}$ is an antisymmetric tensor therefore we always have:
\be
\det(g_{ab}+A_{ab})=\det(g_{ab}-A_{ab})\, .
\ee
So we able to find:
\begin{align}
    \det(g_{ab}+A_{ab})&=\sqrt{\det(g_{ab}+A_{ab})\det(g_{ab}-A_{ab})}\notag\\&=\sqrt{ \det g_{ab}}\sqrt{ det(g_{ab}-A_{ac}g^{cd}A_{db})} \notag\\
    &=\sqrt{ \det g_{ab}}\sqrt{ \det \tilde{g}_{ab}}\,,  \label{ef:met} 
\end{align}
where we introduce \textit{effective metric} as : $\tilde{g}_{ab}=g_{ab}-A_{ac}g^{cd}A_{db}$ .\\
For example, from Eq.\eqref{eq:gauge} and Eq.\eqref{eq:bfield} we would have
\begin{align}\label{eff:metr}
\tilde{g}_{++}&=g_{++}+g^{\a_2 \a_2} B^{2}_{+\a_2}+g^{\a_3 \a_3} B^{2}_{+\a_3}+(2\pi \a')^2 F^{2}_{+x} g^{xx},\\
\tilde{g}_{--}&=g_{--}+g^{\a_2 \a_2} B^{2}_{-\a_2}+g^{\a_3 \a_3} B^{2}_{-\a_3}+(2\pi \a')^2 F^{2}_{-x} g^{xx},\\
\tilde{g}_{rr}&=g_{rr}+(2\pi \a')^2 F^{2}_{rx}g^{rr},\\
\tilde{g}_{+-}&=g_{+-}+B_{+\a_2}B_{-\a_2} g^{\a_2 \a_2}+B_{+\a_3}B_{-\a_3} g^{\a_3 \a_3}+ (2\pi \a')^2F_{+x}F_{-x}g^{xx}.
\end{align}
The horizon of $\tilde{g}_{ab}$ will meet the reality constraint on DBI action i.e., $r_*$  in Eq.\eqref{eq:con2}. So we could assign a geometric meaning to the external electric field. For a more detailed discussion please see~\cite{Kim:2011zd}.
%

\end{document}